\documentclass[a4paper]{jpconf}

\usepackage{graphicx}

\begin{document}

\title{Generalising the logistic map through the $q$-product} 

\author{R W S Pessoa, E P Borges}

\address{Escola Polit\'ecnica, Universidade Federal da Bahia, 
         Rua Aristides Novis 2, Salvador, Bahia, Brazil}

\ead{robsonpessoa2007@gmail.com,ernesto@ufba.br}

\begin{abstract}
We investigate a generalisation of the logistic map as
$ x_{n+1}=1-ax_{n}\otimes_{q_{map}} x_{n}$ 
($-1 \le x_{n} \le 1$, $0<a\le2$)
where $\otimes_q$ stands for a generalisation of the
ordinary product, known as $q$-product 
[Borges, E.P. Physica A {\bf 340}, 95 (2004)]. 
The usual product, and consequently the usual logistic map, 
is recovered in the limit $q\to 1$,
The tent map is also a particular case for $q_{map}\to\infty$.
The generalisation of this (and others) algebraic 
operator has been widely used within nonextensive statistical
mechanics context 
(see C.~Tsallis, {\em Introduction to Nonextensive Statistical Mechanics}, 
Springer, NY, 2009).
We focus the analysis for $q_{map}>1$ at the edge of chaos, 
particularly at the first critical point $a_c$, that depends on the value of 
$q_{map}$.
Bifurcation diagrams, sensitivity to initial conditions, 
fractal dimension and rate of entropy growth 
are evaluated at $a_c(q_{map})$, and
connections with nonextensive statistical mechanics are explored.
\end{abstract}


\section{Introduction}
\label{sec:intro}
Low-dimensional non-linear maps represent paradigmatic models
 in the analysis of dynamic systems. The discrete time
 evolution and the small number of relatively simple equations
 make their treatment easy, 
 without losing  
 the richness of the behaviour, exhibiting order, chaos and a
 well defined transition between them 
 (see, for example, \cite{hilborn,Tsallis1997}).

Strongly chaotic systems are of special interest for 
 statistical mechanics, 
 once they feature well known characteristics:
 exponential sensitivity to the initial conditions,
 ergodicity,
 exponential relaxation to the equilibrium state,
 gaussian distributions \cite{BeckSchlogl}.

In-between ordered systems (with negative Lyapunov exponent)
 and (strongly) chaotic systems (with positive Lyapunov exponent)
 there are those with zero maximal Lyapunov exponent.
 These systems are characterised by
 power-law sensitivity to initial conditions, instead of 
 the exponential sensitivity, and thus are considered as
 weak chaotic systems.
 This change in the dynamics
 may lead to break of ergodicity,
 non-exponential relaxation to equilibrium 
 and/or non-gaussian distributions.
 These behaviours are usually expected to be found
 in systems that are described by nonextensive statistical
 mechanics \cite{ct:1988,ct:springer}.
 Some low dimensional maps, {\em e.g.} the logistic map,
 also exhibit weak chaoticity at the edge of chaos, 
 and hence the interest in studying them to better
 understand nonextensivity.

Power-law like sensitivity to initial conditions 
 and power-law like relaxation to the attractor 
 (more precisely a $q$-exponential law) 
 have already been found in logistic-like maps 
 \cite{Tsallis1997, Moura}.
 $q$-exponential function
 ($e_q^x\equiv [1+(1-q)x]_+^{1/(1-q)}$, the subscript $_+$ is explained 
 in the following)
 appear within nonextensive statistical mechanics 
 and it generalises the usual exponential function
 (recovered as $q\to 1$).
 It is asymptotically a power-law 
 (for $q>1$ and $x<0$ or $q<1$ and $x>0$).
 Sensitivity to initial conditions of the logistic
 map at the edge of chaos is identified to a
 $q$-exponential, with a specific value of the
 parameter $q$, denoted as $q_{sen}$.
The rate of entropy growth  $S_{q}/t$ 
($S_q$ is the nonextensive entropy, defined later by Eq.~(\ref{eq:Sq}),
and $t$ is time)
is another parameter usually evaluated in maps. It must be finite at
the macroscopic limit, and there is one special
value of $q$ denoted $q_{ent}$ (from {\em entropy})
that makes $S_{q_{ent}}/t$ finite. At the edge of chaos,
$q_{ent}\ne 1$.
It is numerically verified that 
$q_{sen}=q_{ent}$ (see \cite{ct:springer} and references therein).
Relaxation of the logistic map to the attractor
 at the edge of chaos
 also follows a $q$-exponential behaviour, with a
 different and specific value of the parameter $q$,
 denoted $q_{rel}$. 
The relation between $q_{ent}\le 1$ and $q_{rel}\ge 1$ 
 plays a central role in the foundations of nonextensive
 statistical mechanics. For completely chaotic systems,
 these values collapse to $q_{ent}=q_{rel}=1$  
 (See \cite{ct:springer} for details).

\begin{figure}
 \begin{minipage}[t]{0.45\linewidth}  
 \includegraphics[width=\linewidth]{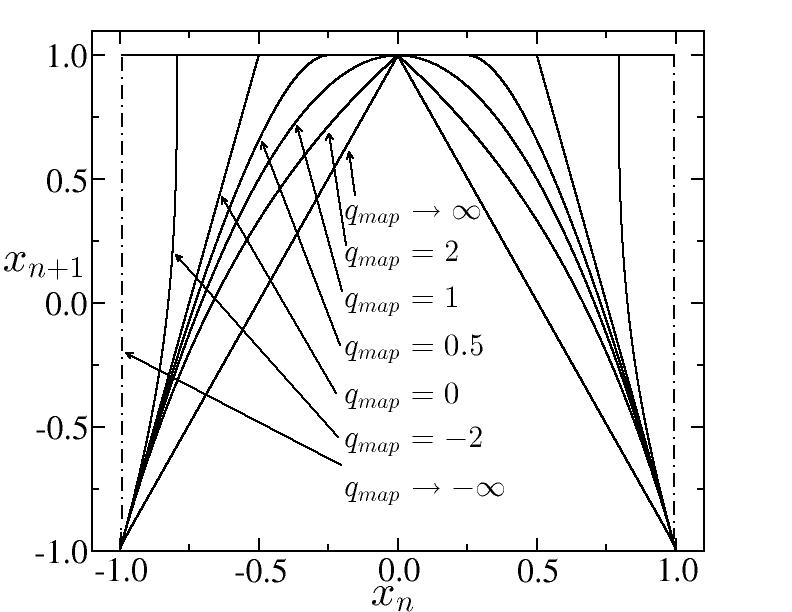}
 \caption{{$x_{n+1}$} as a function of {$x_n$} 
          for the {$q$}-logistic map (Eq.~(\protect\ref{map}) with $a=2$).
          The usual parabolic behaviour is recovered at {$q_{map}=1$}.
          Tent map is found at {$q\to\infty$} (with the peak parameter 
          set to unit).
          As $q_{map}$ departs from 1 towards $-\infty$ ($q_{map}<1$)
          the cut-off condition in Eq.~(\protect\ref{qproduct})
          yields an increasing region in which $x_{n+1}=1$. 
         For {$q_{map}\to-\infty$}, $x_{n>0}$ alternates between $1$ and $(1-a)$ 
          {$\forall x_0 \in [-1,1]$}.
          For $q_{map}>1$ the map is discontinuous at $x=0$ and
          for $q_{map}<1$ the discontinuity is at $|x|=1/2^{1/(1-q)}$.}
 \label{fig:xn1vsxn}
\end{minipage}\hfill
\begin{minipage}[t]{0.45\linewidth}
\includegraphics[width=\linewidth]{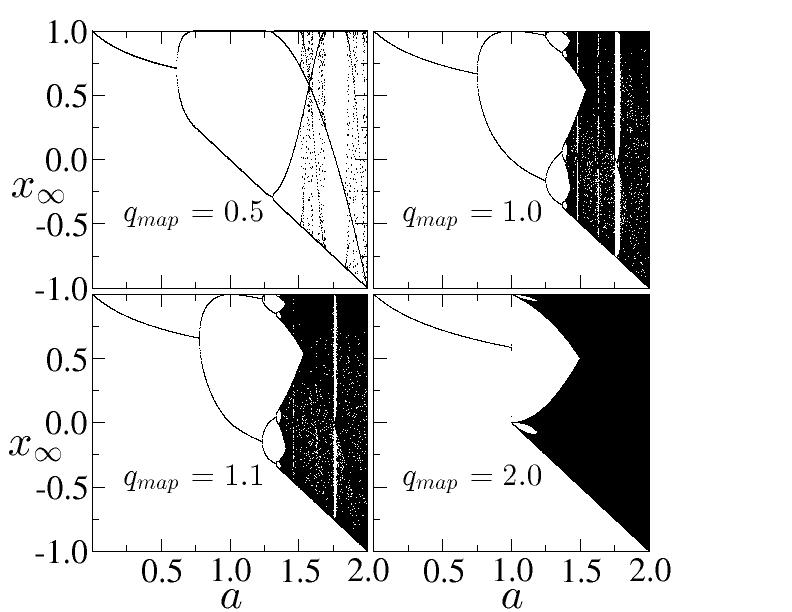}
 \caption{Bifurcation diagrams for different values of 
           {$q_{map}$} (indicated).
           Windows of order inside chaos vanishes as 
 	  {$q_{map}\to 2$}.
	  In this paper we explore $q_{map}>1$ but $q_{map}=0.5$
	  is shown as an instance just to give an idea of the scenario 
	  for $q_{map}<1$: the regions of chaos become narrower and
	  the regions of order become dominant.
	  }
  \label{fig:dbif}
\end{minipage}
\end{figure}

Nonextensive statistical mechanics has lead to developments
 in many related areas, including generalised algebras
 \cite{Wang2003,Borges2004}. These works have introduced 
 generalised algebraic operators, and here we are particularly
 interested in the $q$-product\footnote{The $q$-product has been also used
 in the generalisation of Gauss's law of errors \cite{Suyari2005}, 
 in the formulation of the $q$-Fourier transform, 
 and in the generalisation of the central limit theorem \cite{Umarov2008}.}, 
 defined as  
 \begin{eqnarray}
  \label{qproduct}
  x \otimes_q y \equiv {\rm sign}(x) {\rm sign}(y) \left[|x|^{1-q}+|y|^{1-q}-1\right]^{\frac{1}{1-q}}_+
 \end{eqnarray}
 where the symbol $[A]_+$ means that $[A]_+=A$ if $A>0$ and $[A]_+=0$
 if $A\le 0$ (known as cut-off condition, 
 $[A]_+\equiv \max\{0,A\}$, for short). 
 The limit $q\to 1$ recovers the usual product
 ($x\otimes_1 y = xy$).
Our work consists in generalising the logistic map as
 \begin{eqnarray}
 \label{map}
  \begin{array}{rcl}
  x_{n+1}&=&1-a \left(x_{n}\otimes_{q_{map}} x_{n} \right) \\
   &=& 1 - a[2|x_{n}|^{1-q_{map}}-1]^{\frac{1}{1-q_{map}}}_{+}
  \end{array}
 \end{eqnarray}
($-1 \le x_n \le 1$, $0<a\le2$).
The cut-off condition implies that $[2|x_{n}|^{1-q_{map}}-1]_+ =0$, and thus
$x_{n+1}=1$, if $|x_n|\le 1/2^{1/(1-q_{map})}$.
This $q$-logistic map recovers the usual logistic map for $q_{map}=1$,
 and also the tent map for $q_{map}\rightarrow+\infty$
 (the tent map properly shifted as $x_{n+1}=1-a|x_n|$).
 At the limit $q_{map}\rightarrow-\infty$, it becomes 
 $x_{n+1}=1$ for $-1<x_n<1$ and $x_{n+1}=1-a$ for $x_n=\pm 1$.
 Some details regarding these limits are sketched in \ref{app:limits}. 
 Figure~\ref{fig:xn1vsxn} shows one iteration of the map.

The $z$-logistic map
($x_{n+1}=1-a|x_n|^z$, $z>1$, $0<a\le2$) 
\cite{Costa1997,Borges2002,Crisogno,Tirnakli2006},
is another generalisation of the logistic map for a general power $z>1$
and holds some similarity to the present $q$-logistic map.

Bifurcation diagrams for different values of $q_{map}$ 
are shown in  Fig.~\ref{fig:dbif}.
As $q_{map}$ goes from 1 (the usual logistic map) to 2, 
the value of the parameter $a$ for the first bifurcation
goes from $a=0.75$ to $a=1$ (it approaches 1 from the left).
Also the value of the parameter for the accumulation of bifurcations 
goes from $a_c=1.401155189092\dots$ to $a_c=1$
(it approaches 1 from the right). 
It means that the period doubling cascade becomes narrower until 
it eventually disappears at $q_{map}=2$.
Similar behaviour also happens with the other 
 windows of order inside chaos (where there is tangent bifurcation):
 they get narrower as $q_{map}$ increases, 
 and eventually disappear for $q_{map}=2$.
 Complete chaos is preserved at $a=2$, $\forall q_{map}$.
The Schwarzian derivative for the $q$-logistic map is given by 
(see \ref{app:schwarzian})

\begin{eqnarray}
 \label{eq:schwarzian}
  (Sf)(x) =\left(q_{map}-\frac{1}{2}q_{map}^2\right)
  \frac{1}{|x|^2}
  \left(1-\frac{4}{4-4|x|^{q_{map}-1}+|x|^{2(q_{map}-1)}}\right).
\end{eqnarray}
This expression is negative in the interval $0<q_{map}<2$, 
and positive for $q_{map}>2$ and for $q_{map}<0$
($SD(f(x))=0$ for $q_{map}=0$ and for $q_{map}=2$).
This means that the route to chaos for $0 < q_{map}<2$ is by
period doubling bifurcation.


\section{Sensitivity to initial conditions}

\begin{figure}
\begin{center}           
\includegraphics[width=0.3\linewidth]{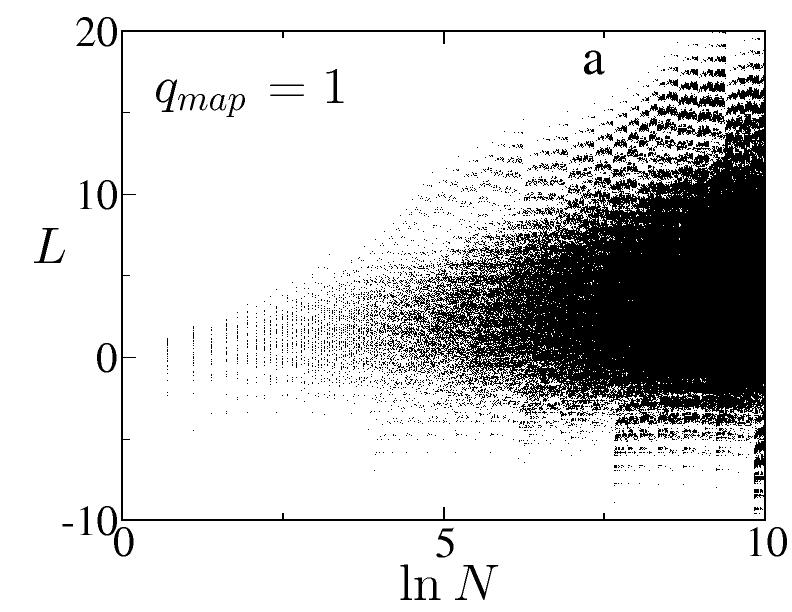}  
\includegraphics[width=0.3\linewidth]{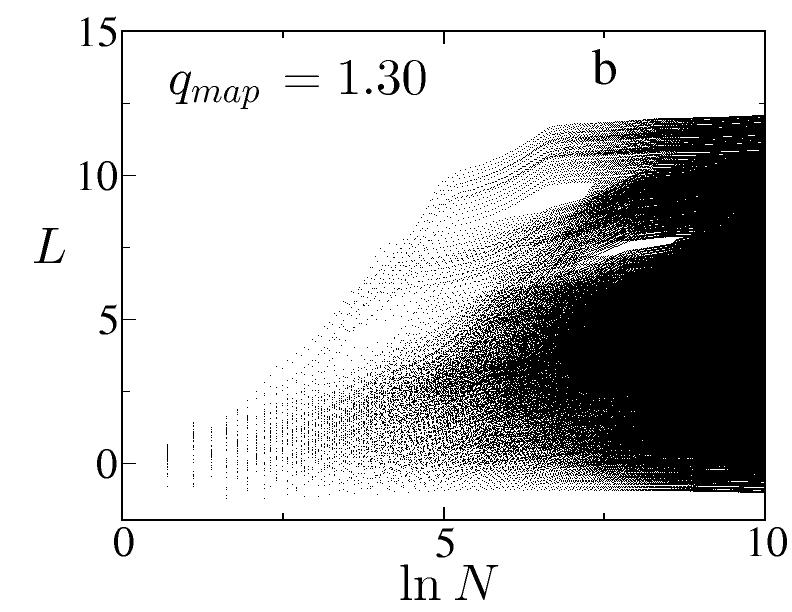}
\includegraphics[width=0.3\linewidth]{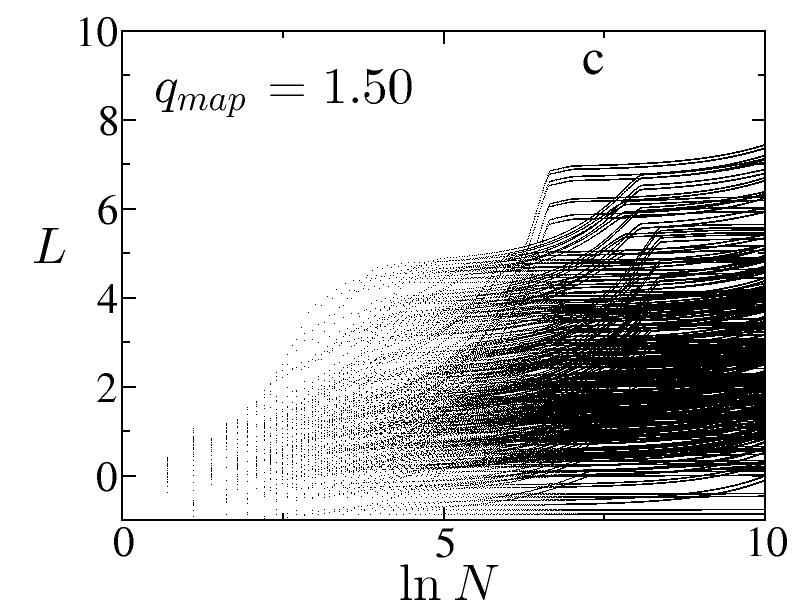} 
\includegraphics[width=0.3\linewidth]{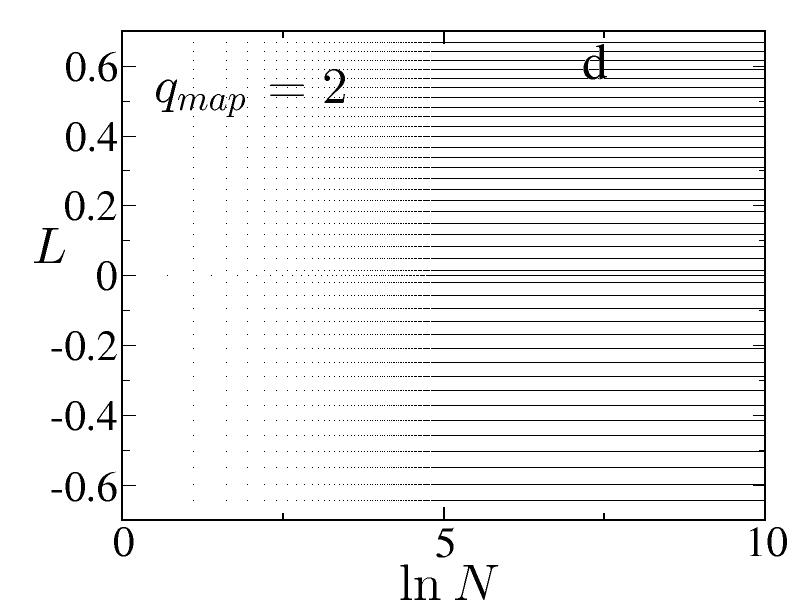}  
\includegraphics[width=0.3\linewidth]{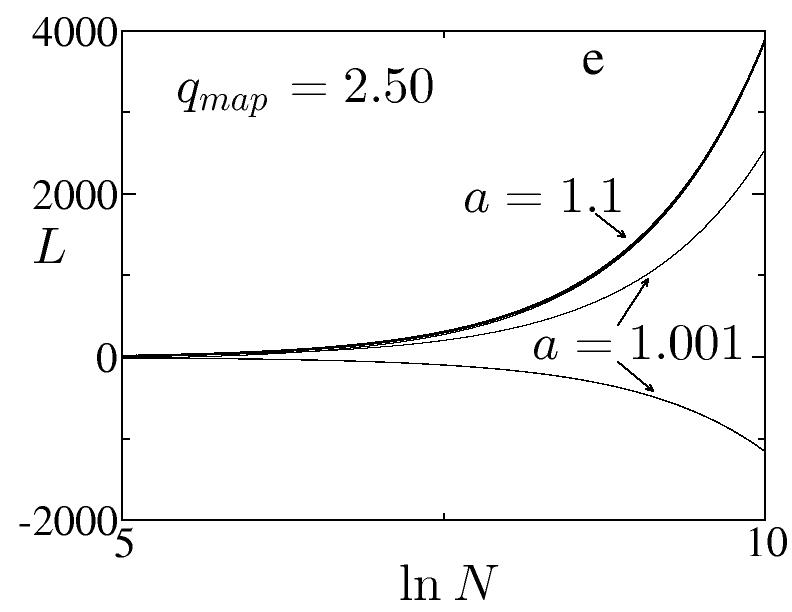}
\includegraphics[width=0.3\linewidth]{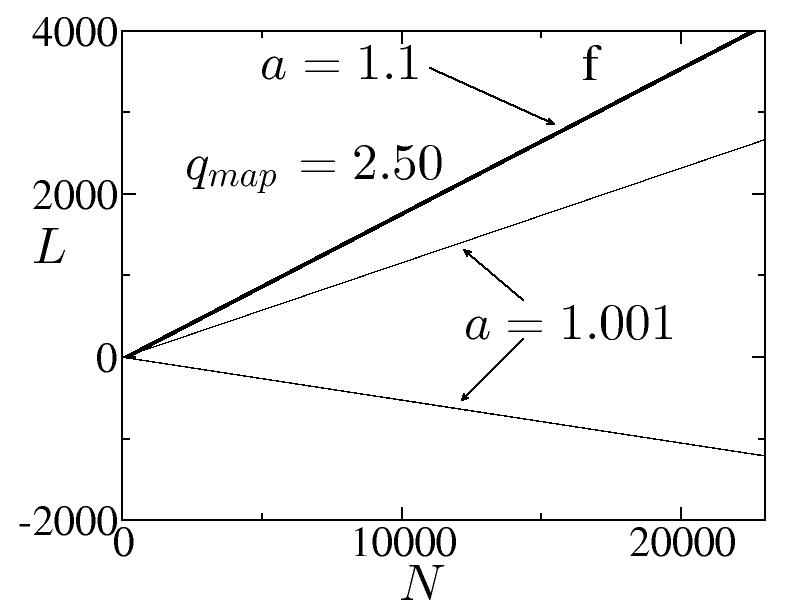}
\caption{Sensitivity to initial conditions as defined by
         variable $L$ (Eq.~(\protect\ref{eq:Lqmap})) for
	 different values of $q_{map}$ (indicated).
         40 initial conditions are used.
	 For $q_{map}=2.5$ (figures e and f) two values
	 of the control parameter $a$ are presented. 
	 It can be seen that it is possible
	 to have coexistence of attractors for a certain ensemble
	 of initial conditions (for $a=1.001$), 
	 one with $L>0$ and the other with $L<0$.
	 The same value of $q_{map}=2.5$ and with $a=1.1$, the initial
	 conditions always lead to $L>0$.
	 Fig.~f shows linear dependence of $L$ with $N$.
	 Slopes (in absolute values) for $a=1.001$ for $L>0$ and $L<0$
	 differ.
	 }
\label{fig:trajetor}
\end{center}
\end{figure}

The maximal Lyapunov exponent may be evaluated at the edge of chaos
according to (see, for instance, \cite{BeckSchlogl})
 \begin{eqnarray}
 \lambda_{max} = \lim_{N\to\infty}\frac{1}{N}\sum^{N-1}_{i=0}\ln{|f'(x_i)|}
 \label{eq:Lyapunov}
 \end{eqnarray}
where $f'(x)$ is the derivative of the $q$-logistic map,
 \begin{eqnarray}
  f{'}(x)=-|x|' 2a|x|^{-q_{map}}\left[2|x|^{1-q_{map}}-1\right]_+^{\frac{q_{map}}{1-q_{map}}}.
 \label{dqmap}
 \end{eqnarray}
$|x|'$ is the derivative of the absolute value function:
$|x|'=1$ for $x>0$ and $|x|'=-1$ for $x<0$. 
For values of the control parameter $a$ different from critical
ones, the sensitivity to initial conditions are characterised
by exponential divergence at regions of chaos 
and exponential decay at regions of order, {\em i.e.},
positive or negative Lyapunov exponent $\lambda_1$ 
(subscript $1$ will be clear in the following) in 
\begin{eqnarray}
\xi(t)= \lim_{\Delta x(0)\to 0} \frac{\Delta x(t)}{\Delta x(0)} 
= e^{\lambda_1 t} .
\label{eq:lambda1}
\end{eqnarray}
At the edge of chaos it was proposed that the divergence
follows an asymptotic power-law 
(in fact a $q$-exponential law) \cite{Tsallis1997}
characterising a slow dynamics,
 \begin{eqnarray}
 \xi(t) = e_{q_{sen}}^{\lambda_{q_{sen}} t} =
 (1+(1-q_{sen})\lambda_{q_{sen}} t)^{\frac{1}{1-q_{sen}}},
 \label{eq:lambqL}
 \end{eqnarray}
where $\Delta x(0)$ represents the distance between two
neighbouring initial conditions and
$sen$ stands for sensitivity to initial conditions
($q_{sen} \le 1$). 
Eq.~(\ref{eq:lambda1}) is recovered at $q_{sen}=1$ (see \ref{app:limits}) 
and this is the reason for the subscript $1$ in $\lambda_1$, 
Eq.~(\ref{eq:lambda1}).
As a graphical representation of the sensitivity,
it can be defined the variable $L$ as \cite{Tsallis1997}:
 \begin{eqnarray}
 L = \sum^{N-1}_{i=0}\ln{|f'(x_i)|}.
 \label{eq:Lqmap}
 \end{eqnarray}
Figure~\ref{fig:trajetor} shows five instances of $L$ vs. $\ln N$.
For $q_{map}=2$ (Fig.~\ref{fig:trajetor}d) 
the dependence of $L$ on $N$ is very slow
and cannot be seen up to $N=2^{15}$ (the upper limit of the figures).
For $q_{map}>2$ it is possible to have coexistence of attractors
according to the control parameter $a$: depending on the initial
conditions, $L$ may be increasingly positive or decreasingly
negative. Fig \ref{fig:trajetor}f shows that there is a
linear dependence of $L$ on $N$ 
but with different slopes for the cases $L>0$ and $L<0$.  
For values of $a>1.07499\ldots$, $L<0$ is never exhibited.
Coexistence of attractors was also found in \cite{jaganathan-sinha}
for a different deformation of the logistic map
(the authors also call their deformation as $q$-logistic map).

\begin{figure}
\begin{center}           
\includegraphics[width=0.3\linewidth]{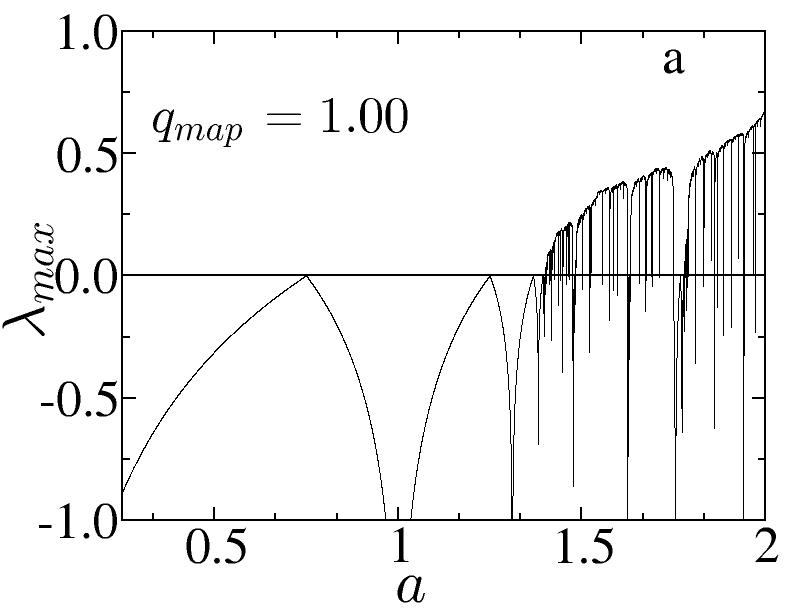}
\includegraphics[width=0.3\linewidth]{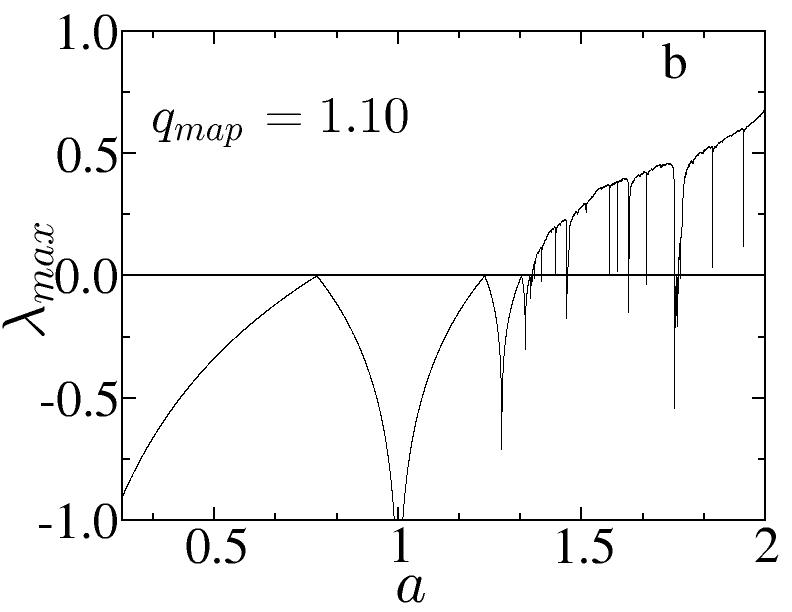}
\includegraphics[width=0.3\linewidth]{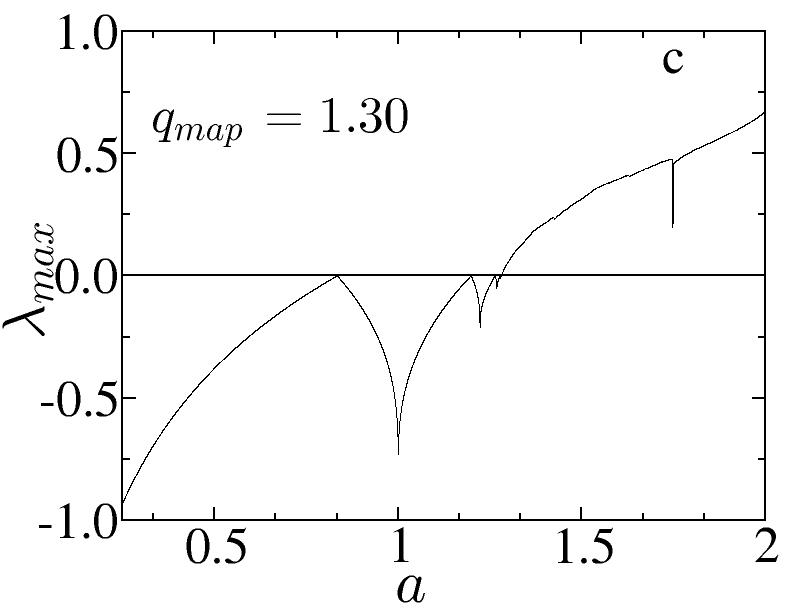}
\includegraphics[width=0.3\linewidth]{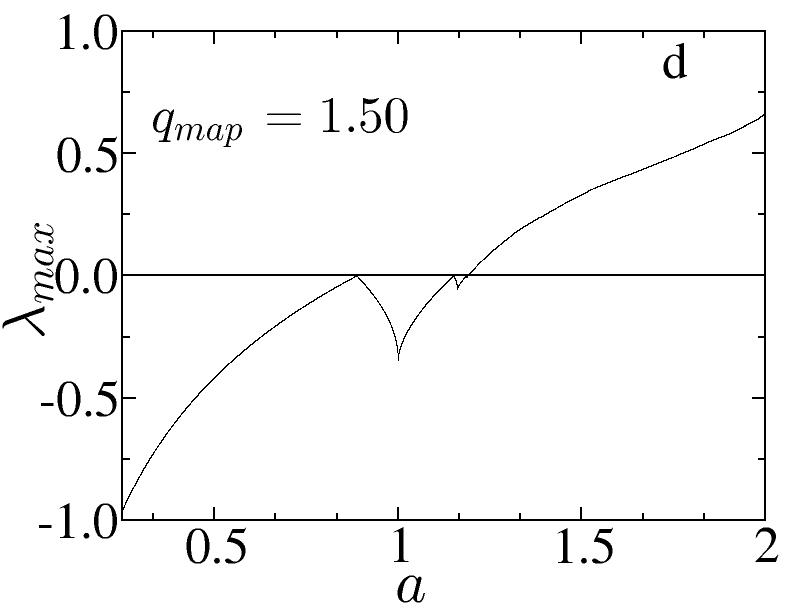}
\includegraphics[width=0.3\linewidth]{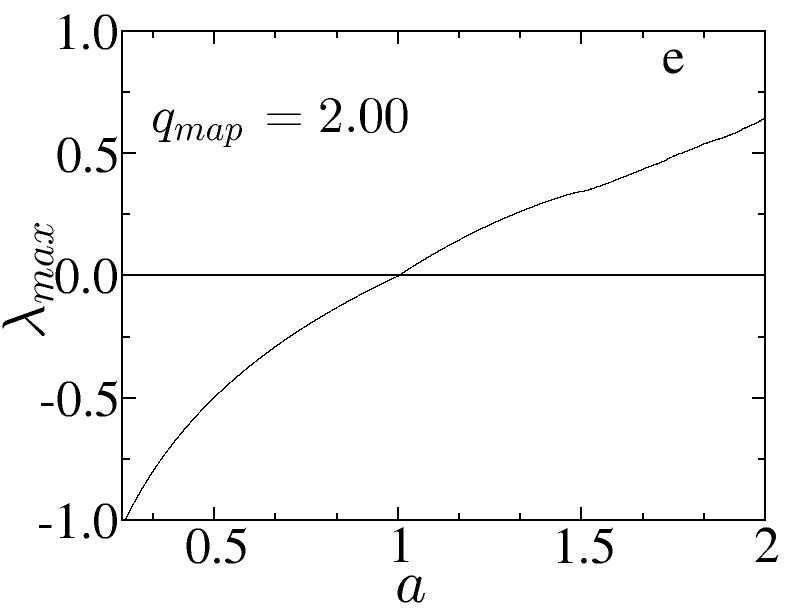}
\includegraphics[width=0.3\linewidth]{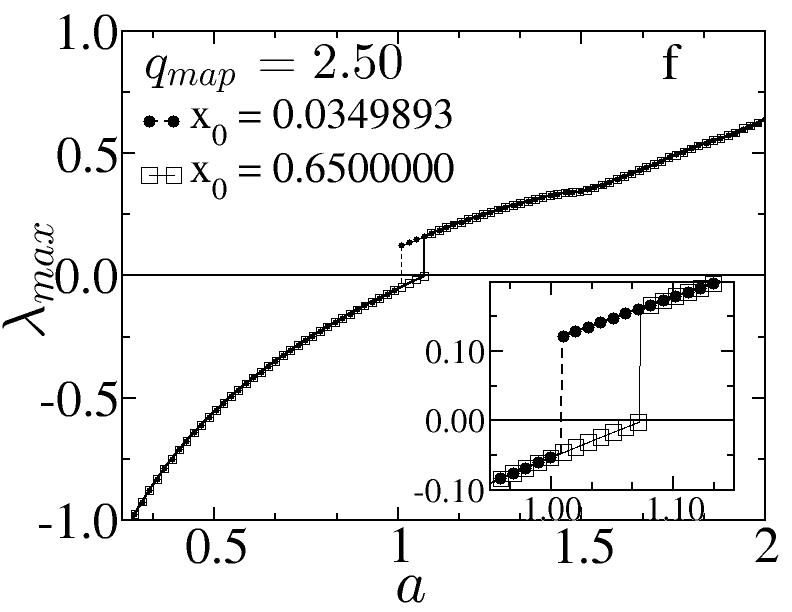}
\caption {Lyapunov exponent as a function of the control parameter $a$
          for different values of $q_{map}$ (indicated).
	  Up to the time limit used in Eq.~(\protect\ref{eq:Lyapunov}),
	  the almost vertical line in Fig.~c at $a=1.7477$ does not crosses
	  zero, but it does present negative values if the calculation 
	  is done with a greater precision (it was used $t=2^{18}$ and
	  $\Delta a = 2 \times 10^{-4}$).
	  Fig.~f exhibits coexistence of attractors for two different
	  initial conditions. It's inset is an amplification of the
	  region of coexistence of attractors for two initial conditions
	  (indicated as full circles and open squares).
	  }
\label{fig:lyapunov}
\end{center}
\end{figure}

Lyapunov exponents are displayed in Fig.~\ref{fig:lyapunov}
 showing transitions from order to chaos. The figures show
 that these transitions become sparse as $q_{map}$ departures 
 from unit, and for $q_{map}=2$ there is only one transition
 (robust chaos)\cite{grebogi1998}.
 Fig.~\ref{fig:lyapunov}f shows coexistence of attractors for $q_{map}=2.5$.

\begin{figure}
\begin{center}
\includegraphics[width=0.5\linewidth]{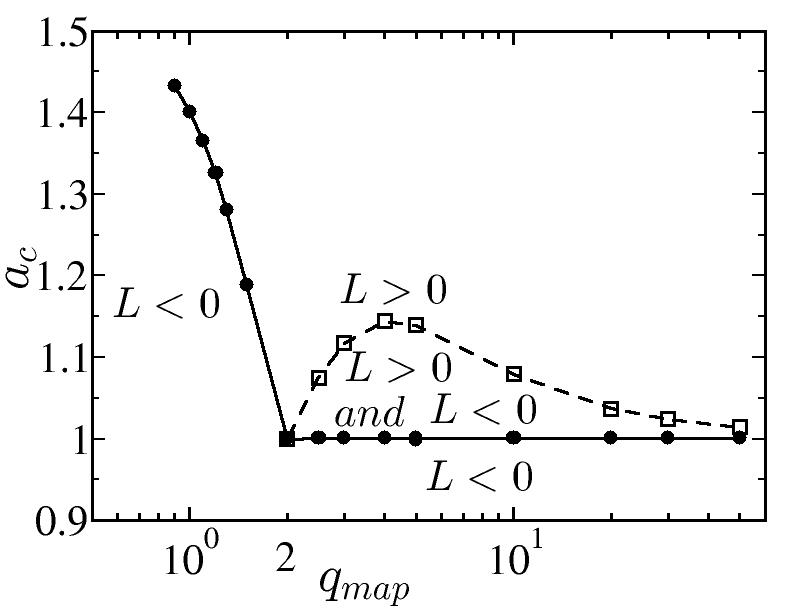}
\caption{Dependence of the first point of accumulation
         of period doubling bifurcation $a_c$ on $q_{map}$.
	 For $q_{map}>2$ there is a region that exhibits
	 coexistence of attractors depending on the initial conditions.
	 Values of $L$ for this figure were calculated without transient time
	 and final time $t_{end}=2^{18}$, 
	 and increments for the control parameter $\Delta{a}=10^{-7}$.
	 Critical point is taken as that for which
         $0<\lambda_{max}<5 \times 10^{-5}$.
         }
\label{fig:retaproducta}
\end{center}
\end{figure}

Figure~\ref{fig:retaproducta} illustrates the first point of 
 accumulation of period doubling bifurcation $a_c$ as a function 
 of $q_{map}$. 
 For $1<q_{map}<2$, the behaviour is ordinary in the sense 
 that for $a<a_c$, $L<0$, and for $a$ slightly greater than $a_c$, $L>0$.
 For $q_{map}>2$ a different behaviour appears:
 order is found for $a<1$ (region below the solid line),
 while chaos is found above the dashed line.
 The region in-between presents coexistence of
 ordered and chaotic behaviours, depending on the
 initial conditions (see figures \ref{fig:trajetor}e, \ref{fig:trajetor}f
 and \ref{fig:lyapunov}f).

Table \ref{aggiungi} shows the range of the critical points $a_c$ 
(first point of accumulation of bifurcations) for different values
of $q_{map}$. The value of $a_c$ is between $a_{-}$ and 
$a_{+}$.  The fourth column shows adopted value for $a_c$.
For the evaluation of $a_c$ listed in Table \ref{aggiungi}, 
the Lyapunov exponents were calculated with a transient time of 
$t_{trans}=2^{23}$ and a final time of $t_{end}=2^{23}$
(the values of the Table are more accurate than those of 
Fig.~\ref{fig:retaproducta}).
Initial condition was fixed in $x_0=0.65$ for all cases.

 \begin{table}
\caption{Critical points}
\label{aggiungi}\centering %
\begin{tabular}{cr@{.}lr@{.}lr@{.}l}
\br
$q_{map}$ & \multicolumn{2}{c}{$a_{-}$} & \multicolumn{2}{c}{$a_{+}$} & \multicolumn{2}{c}{$a_c$}          \\
\mr
1.00      & 1&40115518       & 1&40115520       & 1&401155189092      \\
1.01      & 1&3977569        & 1&3977571        & 1&397757026         \\
1.02      & 1&3943177        & 1&3943179        & 1&394317802         \\
1.03      & 1&3908370        & 1&3908372        & 1&390837098         \\
1.04      & 1&3873144        & 1&3873146        & 1&387314512         \\
1.05      & 1&3837496        & 1&3837498        & 1&383749669         \\
1.10      & 1&3652805        & 1&3652807        & 1&365280586         \\
1.20      & 1&3250906        & 1&3250908        & 1&325090670         \\
1.30      & 1&2811360        & 1&2811362        & 1&281136143         \\
1.40      & 1&2353387        & 1&2353389        & 1&235338767         \\
1.45      & 1&2125150        & 1&2125152        & 1&21251512\;\;      \\
1.50      & 1&1900820         & 1&1900822         & 1&1900822 $\quad$ \\
\br
\end{tabular}
\end{table}

\begin{figure}[htb]
 \begin{center}
  \includegraphics[width=0.5\linewidth]{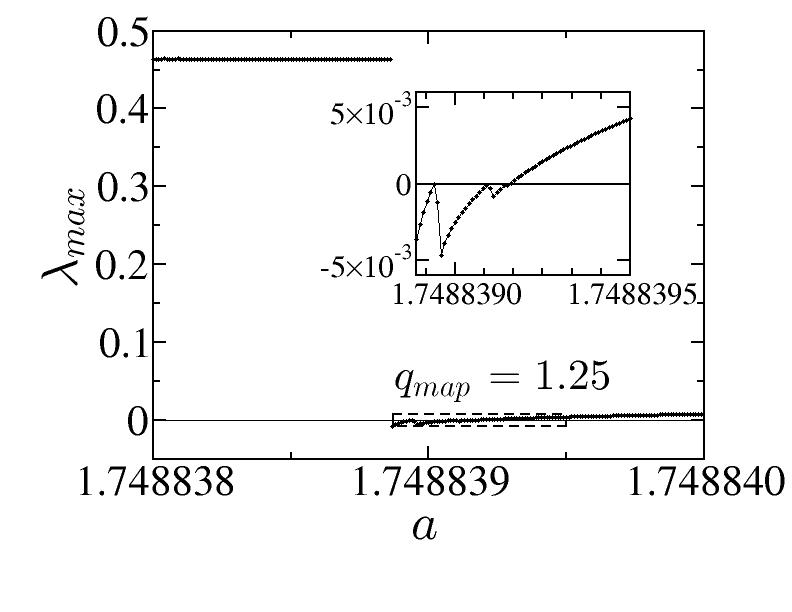}
   \caption{Maximal Lyapunov exponent for $q_{map}=1.25$
   at the vicinity of the largest window, of cycle 3,
   with increments of $\Delta a  = 1.0\times10^{-8}$. 
   It was used a transient time of $t_{trans}=2^{23}$ 
   and the time used to estimate the Lyapunov exponent 
   $t_{end}=2^{23}$ (with lower time intervals the transition
   from negative to positive Lyapunov exponent cannot be seen).
   Inset corresponds to the dashed rectangle in the main panel 
   and shows points of zero Lyapunov exponents that corresponds 
   to period doubling bifurcation.}
   \label{fig:Lyapunovciclo3}
 \end{center}
 \end{figure}

In Fig.~\ref{fig:Lyapunovciclo3} we show the cycle 3 window for
$q_{map}=1.25$. This window of order inside chaos 
(as well as all the others) becomes narrow: in order to identify
the Lyapunov exponent in this instance it was necessary to give
increments of $\Delta a  = 1.0\times10^{-8}$ and $t_{end}=2^{23}$,
with a transient $t_{trans}=2^{23}$. 
Identification of windows of order in this $q$-logistic map is computationally 
time consuming as $q_{map}$ departs from unit.

\section{Entropy production}

\begin{figure}[htb]
\begin{center}
\includegraphics[width=0.5\linewidth]{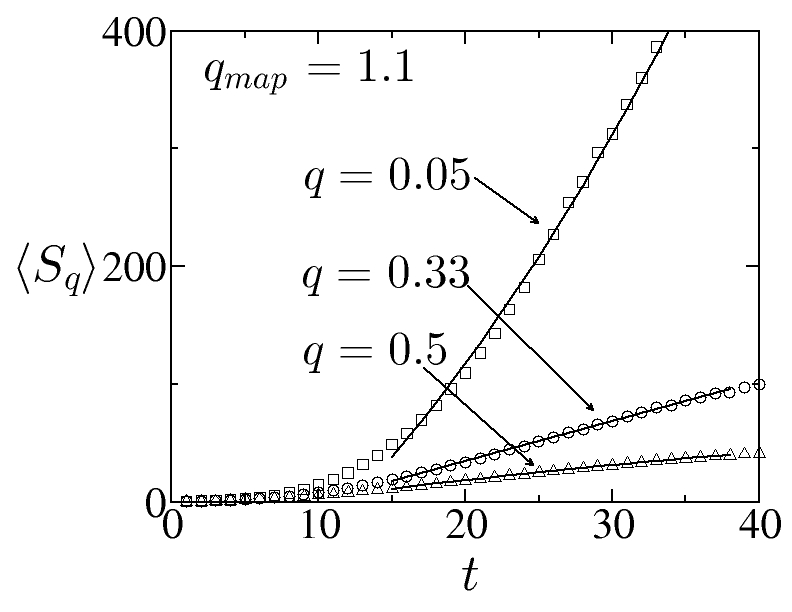}
 \caption{Rate of increase of entropy for $q_{map}=1.1$. We found
 $q_{ent}=0.33$. This value is found by fitting a parabola
 $\langle S_q(t)\rangle = a+bt+ct^2$ for the time interval $15\le t\le 38$
 (the same used in \protect\cite{Latora2000}).
 Initial times are always left out of calculation because the concavity
 is always positive at that region.
 For $q<q_{ent}$ the concavity of the curve is positive
 (squares, $c>0$)
 and thus the rate of growth of entropy diverges at the macroscopic limit,
 which is unphysical. For $q>q_{ent}$, the concavity is negative
 (triangles, $c<0$),
 thus the rate of growth of entropy is zero at the macroscopic limit,
 which is also unphysical. 
 $c=0$ is shown as circles. 
 }
 \label{fig:Sqxtime}
\end{center}
\end{figure}

\begin{figure}[htb]
 \begin{center}
 \includegraphics[width=0.5\linewidth]{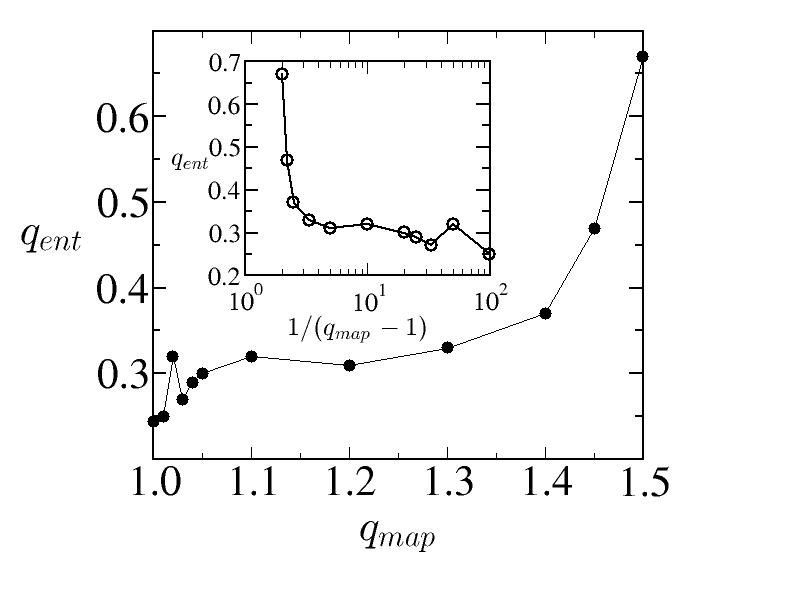}
\caption{$q_{ent}$ as a function of $q_{map}$. 
          Inset shows abscissa in a different scale
	  (note that abscissa is in log scale).
	  The value at $q_{map}=1.02$ may have numerical imprecision.
          Lines are only guide to the eyes.
	  }
 \label{fig:qsenqmap}
 \end{center}
 \end{figure}

Parameter $q_{ent}$ (from {\em entropy}) in Tsallis entropy \cite{ct:1988}
\begin{eqnarray}
 S_q = k\frac{1-\sum_{i=1}^W p_i^q}{q-1}
 \label{eq:Sq}
\end{eqnarray}
($W$ is the number of microstates and $p_i$ is the probability
of 
microstate\footnote{In statistical mechanics, for a given macrostate (that is, 
macroscopically measured state) there are a number $W$ of 
compatible configurations of the constituent elements, or microscopic states
(microstates, for short).
A macroscopic variable is, in fact, a measure of the average 
of all compatible microscopic states.}
$i$; we use $k=1$ without loss of generality)
is estimated according to the method
 developed in \cite{Latora2000}.
 It consists in calculating the rate of increase of entropy,
 which must be {\em finite} for a large (virtually infinite) 
 system%
 \footnote{By large system we mean a macroscopic system. 
 The density variable of a quantity $F$, $F/N$, must remain 
 finite as the number of microscopic constituents $N$ reaches the 
 thermodynamical limit $\lim_{N\to\infty}$.}. 
 The phase space ($x_n\in[-1,1]$) is divided into $W$ cells, 
 with $N$ points (initial conditions) inside one of them.
 As time evolves, Tsallis entropy $S_q$ is calculated for many values of $q$. 
 Calculation is done again with $N$ points inside another initial cell, 
 and this process is repeated for each cell of a certain ensemble
 called ``best initial condition cells''
 (the definition of ``best initial condition cells'' is:
 the integrated number of occupied cells must be greater than a certain
 (arbitrary) threshold, the most visited cells).
 Then it is taken the average of entropy for each time step, 
 and this average is finally plotted against time, for various values of $q$
 (see Fig.~\ref{fig:Sqxtime} for an instance).
 There is only one special value of $q$ for which the increase
 of $S_q$ is linear in the macroscopic limit, that is,
 the production of entropy 
 $\kappa_q=\lim_{t\to\infty} \lim_{W\to\infty} \lim_{N\to\infty} S_q/t$
 remains finite.
 This special value is identified with $q_{ent}$ 
 (see \cite{Latora2000} for details; in that paper, $q_{ent}$ is called $q^*$).
 For values of the control parameter $a$ that corresponds to chaotic behaviour 
 $q_{ent}=1$, so ordinary Boltzmann-Gibbs-Shannon entropy is the
 proper one to be used.
 This is not the case at the edge of chaos ($a=a_c$),
 which we are interested in this work,
 and the value of $q_{ent}$ which yields linear growth
 of entropy (and consequently finite rate at the
 macroscopic limit) is smaller than one
 (for the ordinary logistic map, $q_{map}=1$, at the edge of chaos, 
 $q_{ent}=0.2445...$).
 We applied this procedure to the $q$-logistic map.
The result of this procedure for various values of $q_{map}$ is presented in
Fig.~\ref{fig:qsenqmap}.


\begin{figure}[htb]
\begin{minipage}[t]{0.45\linewidth}
\includegraphics[width=\linewidth]{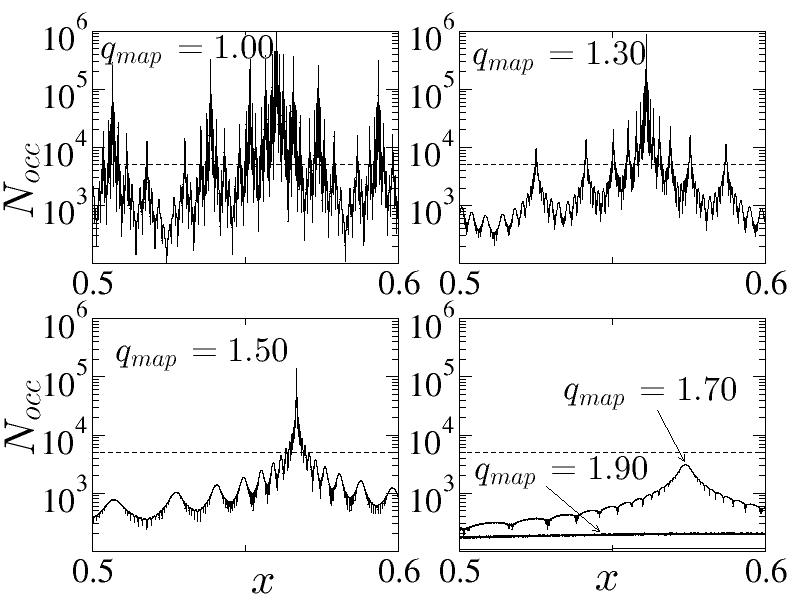}
 \caption{Integrated number of visits per cell for five cases
          ($q_{map} = 1, 1.30, 1.50, 1.70, 1.90$). Phase space is divided
          into $W=10^5$ cells, each one contains $10^6$ points 
	  uniformly distributed. Time evolves up to 50 iterations.
	  Figures show $x\in [0.5,0.6]$ just for better visualisation.
	  }
 \label{fig:casoqrefq}
\end{minipage}\hfill
\begin{minipage}[t]{0.45\linewidth}
\includegraphics[width=\linewidth]{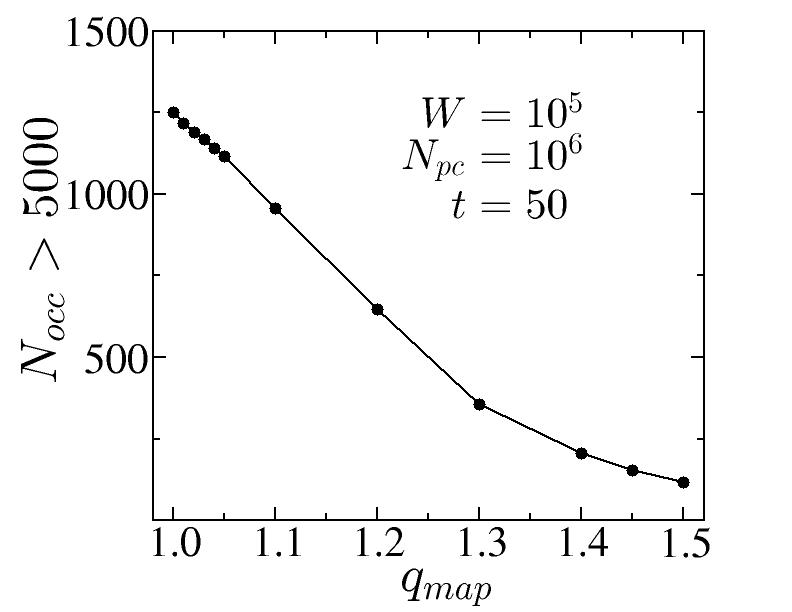}
 \caption{Number of cells with more than 5000 visits in a run of
 50 iterations as a function of $q_{map}$. 
 $N_{pc}=10W$ is the number of initial conditions per cell.
 Dashed horizontal line in Fig.~\ref{fig:casoqrefq}
 indicates the threshold of 5000 visits per cell.
 }
\label{fig:occ}
\end{minipage}
\end{figure}

The integrated number of occupied cells, 
used to identify the best initial condition cells,
 is shown in Fig.~\ref{fig:casoqrefq}.  We see a reduction of
 the number of visited cells as $q_{map}$ goes from one to two
 and a change in the displayed pattern. 
 Fig.~\ref{fig:occ} shows number of cells 
 with visits greater than 5000 as a function of $q_{map}$.

\section{Relaxation to the critical attractor}

\begin{figure}
\begin{center}
\includegraphics[width=0.45\linewidth]{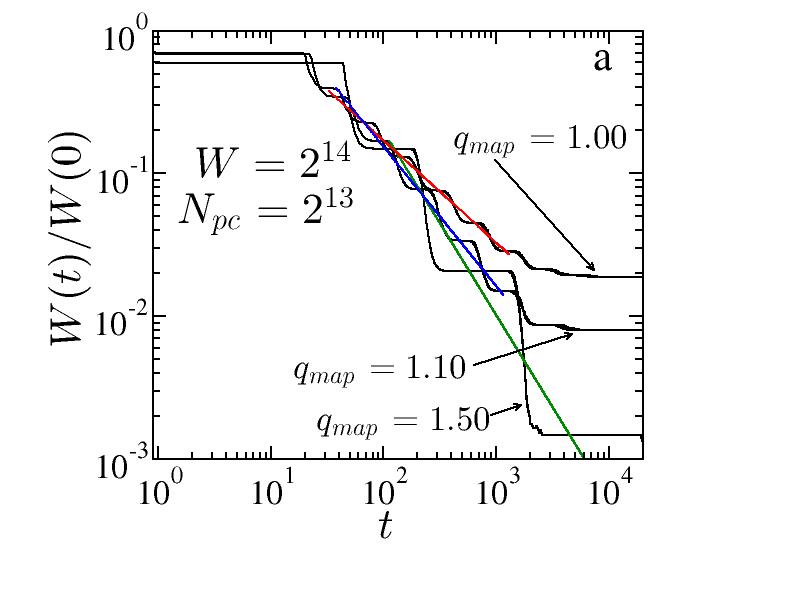} 
\includegraphics[width=0.45\linewidth]{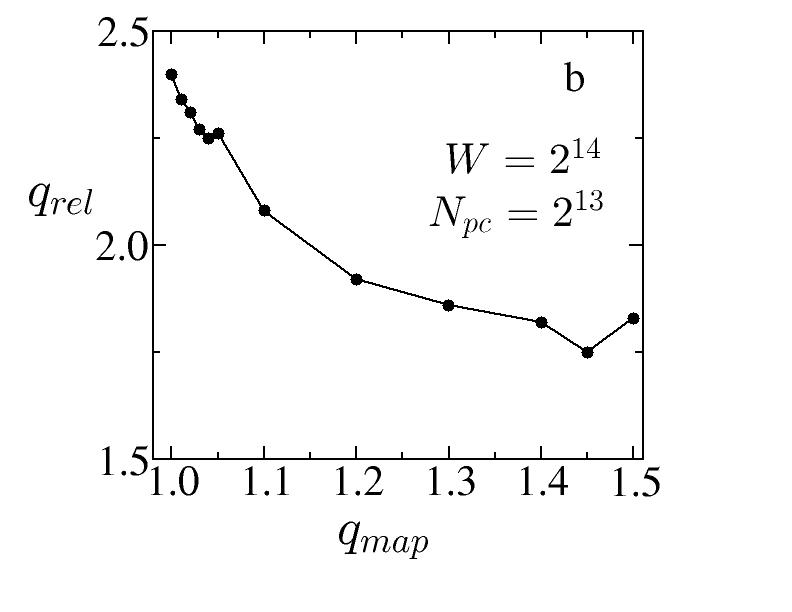}
\caption{Relaxation to the critical attractor at the edge of chaos.
         Fig.~a is a log-log plot of $W_{occ}(t)/W_{occ}(0) \times t$
	 for three different values of $q_{map}$. It can be seen
	 the results of regression of power-laws, whose slopes are
	 identified as $1/(1-q_{rel})$. 
	 Fig.~b shows $q_{rel}$ as a function of $q_{map}$.
	 It was used $W = 2^{14}$ 
	 and number of initial conditions per cell $N_{pc} = 2^{13}$.
         Lines are only guide to the eyes. Fluctuations in the curve of
	 Fig.~b are due to inaccuracy --- the results are
	 very sensitive to small changes in the characteristic parameters.
	 }
\label{fig:qrelqmap}
\end{center}
\end{figure}

Relaxation to the critical attractor was presented at 
\cite{Moura} and basically consists in the division of the phase space
[-1,1] into $W$ cells and take an ensemble of $N$ initial conditions uniformly
distributed in the entire phase space (thus maximal entropy).
Time evolution leads to the decreasing of the number of occupied cells
$W_{occ}(t)$ according to a power-law with log-periodic oscillations
($W_{occ}(0)=W$).
It is supposed \cite{Moura} that the power-law is the asymptotic limit
of a $q$-exponential, 
and the parameter is denoted as $q_{rel}$, for {\em relaxation} ($q_{rel}>1$):
 \begin{eqnarray}
W_{occ}(t) = (1+(1-q_{rel})K_{q_{rel}} t)^{\frac{1}{1-q_{rel}}},
 \label{eq:Wqrel}
 \end{eqnarray}
$K_{q_{rel}}$ is the inverse of a characteristic time.
Figure~\ref{fig:qrelqmap} presents the results for the fraction
of occupied cells $W_{occ}(t)/W_{occ}(0)$ for different values of $q_{map}$
at their corresponding critical points.
Log-periodic oscillations present increasing periods for $q_{map} \to 2$.
The slope in the log-log plot (Fig.~\ref{fig:qrelqmap}a)
at the region of the log-periodic oscillations is used to
estimate $q_{rel}$ as $\mbox{slope}=1/(1-q_{rel})$.
Fig.~\ref{fig:qrelqmap}b presents $q_{rel}$ as a function
of $q_{map}$.

\begin{figure}
 \begin{center}
 \includegraphics[width=0.5\linewidth]{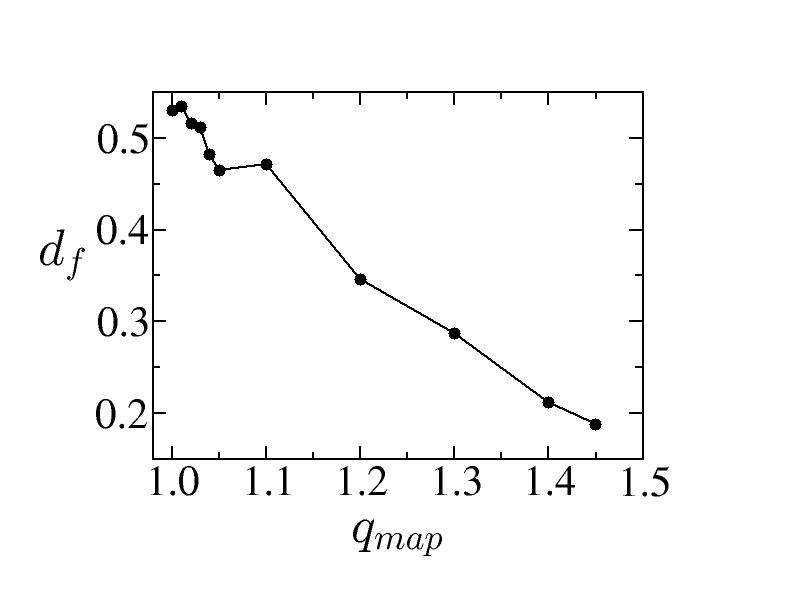}
 \caption{Fractal dimension at the edge of chaos as a function of $q_{map}$.
          For the usual logistic map ($q_{map}=1$), $d_f(a_c)=0.53665$.
          Lines are only guide to the eyes. Fluctuations in the curve
	  are due to inaccuracy --- the method is very sensitive 
	  to small changes in the characteristic parameters.
          }
 \label{fig:fractal}
 \end{center}
\end{figure}

The procedure for estimating $q_{rel}$, that is, the shrinking of
the number of occupied cells, is also used to estimate the fractal dimension 
$d_f$ at the edge of chaos (it is a kind of box counting method).
$W_{occ}(\infty)$ increases with $W_{occ}(0)$ according to 
$W_{occ}(\infty) \propto [W_{occ}(0)]^{d_f}$.
Fractal dimension decreases with $q_{map}$ as shown in Fig.~\ref{fig:fractal}.
%

\section{Final remarks}

The generalisation of the logistic map by means of the $q$-product introduces
some interesting features in its dynamical behaviour.
Our analysis is focused on $q_{map}>1$. In this region, the windows of order
inside chaos become narrower as $q_{map}$ increases until all of them
disappear and the map becomes the tent map.
For $q_{map}<1$ (not analysed in this paper) the opposite behaviour occurs:
the regions of chaos become narrower and order dominates the scenario.
A remarkable feature of the $q$-logistic map is to continuously pass
from a map with a variety of behaviours, such as period doubling
bifurcation, multifractality and power-law like sensibility to
initial conditions at the edge of chaos (the logistic map),
to a robust map (the tent map).
We have calculated the sensitivity to initial conditions, 
the rate of entropy growth, the relaxation to the critical attractor
and the fractal dimension at the edge of chaos.
These methods permit to estimate the parameters $q_{ent}$ and $q_{rel}$
for different values of $q_{map}$. The entropy parameter and
the relaxation parameter are two indices that appear within
the context of nonextensive statistical mechanics and the understanding
of their dependence on the control parameters of the system may lead
to their {\em a priori} determination.
Some other evaluations for this $q$-logistic map remains to be done,
{\em e.g.} the dependence of $q_{rel}$ on the coarse graining $W$ and 
its relation to $q_{sen}$ (as in \cite{Borges2002} for the $z$-logistic map,
and as in \cite{Borges2004c,Borges2004b} for the H\'enon map),
the probability distributions of sums of iterates as in 
\cite{Tirnakli2007,Tirnakli-2009},
multifractality as it was done in \cite{Lyra1998},
and tangent bifurcations.

\section*{Acknowledgements}
This work was partially supported by FAPESB 
(Funda\c{c}\~ao de Amparo \`a Pesquisa do Estado da Bahia).
We thank FESC and GSUMA (research groups of the Institute of Physics of UFBA)
for using their computational resources.


\appendix
\section{Special limits for the $q$-product}
\label{app:limits}

The limit $q\to v$ where $v$ is one of the following $\{-\infty,1,\infty\}$
for the $q$-product $x \otimes_q x$ as it appears in the $q$-logistic map, 
Eq.~(\ref{map}) and also in Eq.~(\ref{qproduct}) with $y=x$, leads to 
$\{0^{0},{1}^{\infty},{\infty}^0\}$ respectively.
The indeterminates can easily be solved by means of a simple trick that is
to rewrite the $q$-product as
\begin{eqnarray}
\label{eq:hqgq}
x \otimes_q x = [2|x|^{1-q}-1]_+^{\frac{1}{1-q}}=h(q)^{g(q)} 
\end{eqnarray}
with $h(q)=[2|x|^{1-q}-1]_+$ 
(note that $h(q)\ge0$ due to the cut-off condition in Eq.~(\ref{qproduct})) 
and $g(q)=\frac{1}{1-q}$.
Of course $h$ is a function of $q$ and $x$ but for now we are not interested
in the dependency on $x$, so we omit it.
We also omit the subscript of $q$ for the sake of brevity.
Then
\begin{eqnarray}
\lim_{q\to v}[2|x|^{1-q}-1]_+^{\frac{1}{1-q}}=
e^{\lim_{q\to v}g(q) \ln h(q)}.
\end{eqnarray}
Straightforward application of L'Hospital rule leads to
(we remind the reader that $|x|\le 1$)
\begin{eqnarray}
\lim_{q\to v}  x \otimes_q x =
\left\{
 \begin{array}{ll}
   \left.
   \begin{array}{cl}
     1, & \mbox{if } |x| = 1 \\
     0, & \mbox{if } |x| < 1 
   \end{array}
   \right\}
  & \mbox{for } v = -\infty, \\
   \begin{array}{ll}
 x^2, &
   \end{array}
 & \mbox{for } v = 1, \\
   \begin{array}{ll}
 |x|, &
   \end{array}
 & \mbox{for } v = \infty.
 \end{array}
\right.
\end{eqnarray}
L'Hospital rule must be applied twice in the case $v=\infty$.
A similar procedure applied to Eq.~(\ref{eq:lambqL}),
now with $h(q)=1+(1-q_{sen})\lambda_{q_{sen}} t$ and $g(q)=\frac{1}{1-q_{sen}}$,
leads to Eq.~(\ref{eq:lambda1}).

\section{Schwarzian derivative for the $q$-logistic map}
\label{app:schwarzian}

The Schwarzian derivative is defined by
\begin{eqnarray}
\label{eq:DS}
(Sf)(x) = \frac{f^{'''}}{f^{'}} - \frac{3}{2}\left(\frac{f^{''}}{f^{'}}\right)^2.
\end{eqnarray}
The function $f(x) = 1-a \left(x\otimes_{q_{map}} x \right)$
represents the $q$-logistic map and it may be written as
$f(x)=1-ah(x)^{g(q)}$ 
with $h(x)$ and $g(q)$ given by Eq.~(\ref{eq:hqgq}) 
--- now we are interested in the dependency of $h$ on $x$.
The three first derivatives of $f(x)$ are given by
\begin{eqnarray}
\begin{array}{lcl}
f^{'}(x) &=& -a\frac{1}{1-q}h^{\frac{q}{1-q}}h^{'}  \\
&&\\
f^{''}(x) &=& -a\frac{1}{1-q}\left(\frac{q}{1-q}h^{\frac{2q-1}{1-q}}\left(h^{'}\right)^{2} + 
h^{\frac{q}{1-q}}h^{''}\right)  \\
&&\\
f^{'''}(x) &=& -a\frac{1}{1-q}\frac{q(2q-1)}{(1-q)^{2}}h^{\frac{3q-2}{1-q}}\left(h^{'}\right)^{3} \\ 
&&-a\frac{1}{1-q}3\frac{q}{1-q}h^{\frac{2q-1}{1-q}}h^{'}h^{''} \\
&&-a\frac{1}{1-q}h^{\frac{q}{1-q}}h^{'''} 
\end{array}
\nonumber
\end{eqnarray}
with
$ h^{'} = |x|' 2(1-q)|x|^{-q} $,
$ h^{''} = -2q(1-q)|x|^{-1-q} $ and
$ h^{'''} = |x|' 2q(1-q)(1+q)|x|^{-2-q}$.
Substitution in Eq.~(\ref{eq:DS}) leads to
%
\begin{eqnarray}
\label{eq:DS-qlogistic}
(Sf)(x) &=& \frac{4q(2q-1)}{|x|^{2q}[2|x|^{1-q}-1]_{+}^2} 
-\frac{6q^2}{|x|^{1+q}[2|x|^{1-q}-1]_{+}}  +  \frac{q(1+q)}{|x|^{2}} \nonumber \\ 
&&-\frac{3}{2}\left(\frac{2q}{|x|^{q}[2|x|^{1-q}-1]_{+}} - \frac{q}{|x|}\right)^{2}
\end{eqnarray}
This expression may be rearranged as in Eq.~(\ref{eq:schwarzian})
that is more convenient to analyse its signal.
The Schwarzian derivative for the $q$-logistic map presents 
a divergence at $x=0$ for $1\le q<2$ and 
a divergence at $x=0$ and at $x=\pm 1/2^{1/(1-q)}$ for $q<1$.
The case $q_{map}=1$ corresponds to the Schwarzian derivative
for the usual logistic map, $(Sf)(x)= -3/(2x^2)$, as can be easily verified.

\section*{References}

\bibliography{PessoaBorges2011}

\end{document}